\documentclass[reprint,bibnotes,amsmath,amssymb,aps,prb,showpacs,floatfix,superscriptaddress]{revtex4-1}

\usepackage{amsmath}
\usepackage{amsfonts}
\usepackage{amssymb}
\usepackage{xcolor}
\usepackage{graphicx}
\usepackage{subfigure}

\usepackage[breaklinks=true,colorlinks,citecolor=blue,linkcolor=blue,urlcolor=blue]{hyperref}

\newcommand{\M}{\mathbb{M}}

\def\be{\begin{equation}}
\def\ee{\end{equation}}

\def\ber{\begin{eqnarray}}
\def\eer{\end{eqnarray}}

\begin{document}

\title{Enhancement of tunneling density of states at a Y junction of  \\ spin-$\frac{1}{2}$ Tomonaga-Luttinger liquid wires}
\author{Sougata Mardanya}
\affiliation{Department of Physics, Indian Institute of Technology Kanpur, Kanpur - 208016, India}
\author{Amit  Agarwal}
\email{amitag@iitk.ac.in}
\affiliation{Department of Physics, Indian Institute of Technology Kanpur, Kanpur - 208016, India}

\date{\today}

\begin{abstract}
We calculate the tunneling density of states (TDOS) in a dissipationless three wire junction of interacting spin-1/2 electrons, and find an anomalous enhancement of the TDOS in the zero bias limit, even for repulsive interactions for several bosonic fixed points. This enhancement is physically related to the reflection of holes from the junction for incident electrons, and it occurs only in the vicinity of the junction ($x < v_{\rm min}/2\omega$ where $v_{\rm min}$ is the minimum of the velocity of charge or spin excitations and $\omega$ is the bias frequency), crossing over to the bulk value which is always suppressed, at larger distances. The TDOS exponent can be directly probed in a STM experiment by measuring the differential tunneling conductance as a function of either the bias voltage or temperature as done in C. Blumenstein {\it et al.},  Nat.  Phys. {\bf 7}, 776 (2011). 
\end{abstract}


\maketitle

\section{Introduction}
 
One dimensional (1D) quantum wires with strong electron-electron (e-e) interactions are described by 
the Tomonaga-Luttinger liquid (TLL) theory 
[\onlinecite{Tomonaga, Mattis, Haldane, Luttinger, delft, Giamarchi}], 
in which the low energy excitations are collective density oscillations. These density oscillations 
or plasmon modes, are fundamentally different from their 2D and 3D counterpart, i.e.,  Landau's 
quasi-particle excitations, which are described very successfully 
by the Fermi liquid theory [\onlinecite{Giovanni}]. This leads to 
unique physics in 1D, such as the spin-charge separation [\onlinecite{Auslaender1, Jompol}], the phenomena of charge 
fractionalization [\onlinecite{Safi, Safi2, Safi3, pham, Steinberg, Kamata}], power law behaviour of the differential tunneling conductance (with bias voltage or temperature) in quantum Hall edge states [\onlinecite{Chang}], carbon nanotubes [\onlinecite{Bockrath}],  tailored quantum wires in semiconductor heterostructures [\onlinecite{Jezouin}] etc.   TLL behaviour has also been observed in  self-aligned Au atomic chains of single-atom width on germanium surface using scanning tunneling spectroscopy and photoemission [\onlinecite{Blumenstein}]. 

\begin{figure}[t]
\includegraphics[width=0.6\linewidth]{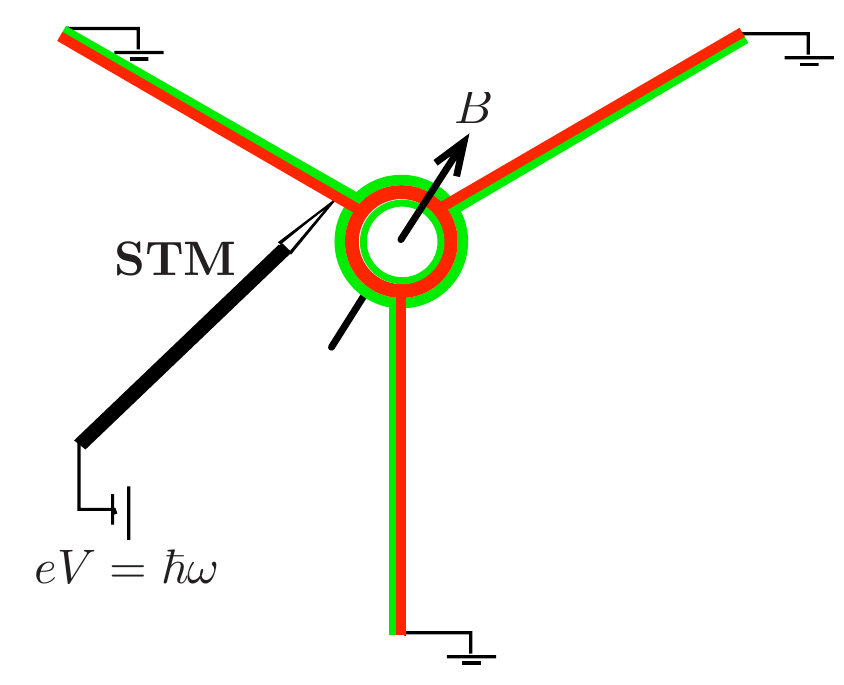}
\caption{Schematic of a Y junction of three spin-1/2 TLL wires, and a STM tip for measuring the TDOS.
The red and green lines indicate the charge and spin sectors respectively, and the inter-wire tunneling region is represented by the circular region. 
\label{fig1}}
\end{figure}

In this paper we explore the local single particle tunneling density of states (TDOS) at a junction of three spin full TLL wires, which can be experimentally probed in a scanning tunneling microscope (STM)  experiment [\onlinecite{flensberg}]. Such Y-junctions 
have already been realized experimentally in crossed single-walled carbon nanotubes  [\onlinecite{fuhrer,terrones}], and have been explored very actively in the literature [\onlinecite{nayak, lal200, chen, chamon1, Meden_prb2005, das2, Giuliano, 
Bellazzini, Hou, agarwal_tdos, abhiram, dasrao, Rahmani, Feldman_PRB2011, 
wolfe1, Hou_prb2012, Agarwal_TRP, agarwal_diss}]. The differential tunneling conductance of a STM tip maintained at a finite bias voltage with respect to the wire [see Fig.~\ref{fig1}] is a direct probe of the locally available electronic states in the TLL, assuming the density of states in the STM tip to be a constant. In a TLL wire the STM current has been shown to vary as  a power of the bias voltage:  
$dI/dV \propto  \rho (\omega) \propto \omega^{\Delta-1}$, with the TDOS exponent $\Delta$ depending on the strength of the e-e interaction strength [\onlinecite{Eggert, Anfuso, Guigou, Pugnetti, Ziani, Dario, Ziani2}]. The TDOS exponent in a spinless two wire junction tuned to the connected (or a single wire without impurity) [\onlinecite{delft,Voit}]  and disconnected  fixed points (single wire with impurity) [\onlinecite{Oreg,Fabrizio,Furusaki,delft}], are known to be $\Delta = (g+g^{-1})/2$ and $\Delta = 1/g$ respectively, where $g$ denotes the TLL interaction parameter. 

Earlier theoretical studies of Y-junctions have used the fermionic language and the weak 
interaction renormalization group (RG) approach [\onlinecite{lal200}], or 
the bosonic and conformal field theory language [\onlinecite{chamon1, das2}], 
or other numerical methods such as the functional RG [\onlinecite{Meden_prb2005}], and were primarily focused on the stability analysis of various 
fixed points of the junction and the associated DC conductivity. 
A detailed analysis of the fixed points of a three wire junction formed from 
spinless interacting electrons was done in 
Ref.~[\onlinecite{chamon1}] which was later extended to include spin-1/2 
electrons giving a much richer phase diagram in the parameter space of spin and charge 
interactions [\onlinecite{Hou}]. 

The tunneling density of states in a three wire Y-junction of spinless electrons was explored in Ref.~[\onlinecite{agarwal_tdos}], 
where the authors highlighted an anomalous enhancement of the TDOS in the vicinity of the junction for several bosonic fixed points. 
Here we consider a more realistic case of a three wire junction formed from spin-1/2 interacting electrons, and obtain an analytic expression for the 
TDOS in the vicinity of the Y-junction. 
This paper is organized as follows: In Sec.~\ref{secII} we describe the 
details of the spin-1/2 TLL  Y-junction and show that both the e-e
interactions in the wire and the current partitioning boundary conditions at the 
junction, can be treated using bosonization with delayed evaluation of the 
boundary conditions [\onlinecite{chamon1}] taking advantage of the spin charge separation in 1D.
In Sec.~\ref{secIII},  we calculate the TDOS of the spin-1/2 TLL Y-junction 
which shows an anomalous enhancement for several fixed points even for repulsive interactions (see Figs.~\ref{fig2} and ~\ref{fig3}).
Finally we summarize our findings in Sec.~\ref{summary}.

\section{Model and Bosonization}
\label{secII}
In this section we review the TLL description of  interacting 1D wires with spin-1/2 electrons which gives rise to spin charge separation in the bulk of the wires. The dissipation-less junction is then described by means of additional tunneling between the wires, which leads to spin and charge current partitioning boundary condition at the junction. The system of three spin-1/2 TLL wires with the junction (Y-junction) is then explored using the delayed evaluation of  boundary conditions first developed in Ref.~[\onlinecite{chamon1}].  We note that the technique of delayed evaluation of boundary conditions is an extension of the plasmon scattering approach first developed in Refs.~[\onlinecite{Safi, Safi2, Safi3}], where the outgoing chiral currents are linearly related to the incoming currents and the conductance is related to the scattering of plasmons.

\subsection{Bosonization of the wires}
Let us consider the Y-junction to be made of three identical, semi-infinite ($x>0$) TLL wires with spin-$\frac{1}{2}$ electrons,  with the junction being located at the origin. For simplicity we consider  all the wires to have the same short ranged e-e interaction strength, and same Fermi velocity for both up-spin and down-spin electrons in all the wires. In each wire, the physical fermionic field for each spin species $\sigma$ can be expressed in terms of slowly varying incoming (left movers) and outgoing (right movers) chiral fields as $\psi_\sigma (x) = e^{i k_{F} x} \psi_\sigma^O + e^{-i k_F x} \psi_\sigma^I $. 
The fermionic field of an outgoing (right-movers) and incoming (left-movers) electron with spin $\sigma$ can be bosonized as,
\begin{eqnarray}\label{1}
\psi_{\sigma}^O(x)&=&\frac{1}{\sqrt{2 \pi \alpha}}F_{O \sigma} ~e^{i2 \pi N_{O \sigma} (x - vt)/L}~e^{- i\varphi_{\sigma}^O (x)}~, \nonumber \\ 
\psi_{\sigma}^I(x)&=&\frac{1}{\sqrt{2 \pi \alpha}}F_{I \sigma} ~e^{i2 \pi N_{I \sigma} (x + vt)/L}~e^{ i\varphi_{\sigma}^I (x)}~,
\end{eqnarray} 
where $\varphi_{\sigma}^a(x,t)$, with $a= O/I$ indicating the propagation direction,  represents the  collective density excitations in 1D.
 In Eq.~\eqref{1}  $F_{a\sigma}$ denotes the Klein factor (or ladder operators) which increase and decrease the corresponding fermion number and satisfy anti-commutation relations among themselves, $\alpha$ is the inverse ultraviolet (short-distance) cut off and $N_{a \sigma}$ counts the number of incoming or outgoing chiral particles of spin $\sigma$ with respect to the filled Fermi sea. The bosonic field operators can be expressed in terms of bosonic creation and annihilation operators as,
\be
\label{3}
\varphi_{\sigma}^a=\sum_{q>0}\frac{1}{\sqrt{n_q}}(b_{\sigma q}^a e^{a iqx}+b^{a \dagger}_{\sigma q} e^{-a iqx})e^{-\alpha|q|/2}~,
\ee
and the bosonic fields satisfy the following equal time commutation relations in a non-interacting theory, $[\varphi_\sigma^a (x), \varphi_{\sigma'}^{a'} (x')] = \pm i \pi \Theta(x-x') \delta_{\sigma \sigma'} \delta_{a a'}$, where the $+$ ($-$) sign arises for $a$ denoting the outgoing (incoming) mode in each wire, and $\Theta(x) $ is the Heaviside step function.   

The corresponding incoming and outgoing charge density and the current fields in each wire are given by
\begin{eqnarray}\label{5}
\rho_{\sigma}^{O}&=&\frac{\partial_{x}\varphi_{\sigma}^{O}}{2\pi}~,~ J_{\sigma}^{O}=-\frac{\partial_{t}\varphi_{\sigma}^{O}}{2\pi}~,\nonumber \\ 
\rho_{\sigma}^{I}&=&-\frac{\partial_{x}\varphi_{\sigma}^{I}}{2\pi}~,~
J_{\sigma}^{I}=\frac{\partial_{t}\varphi_{\sigma}^{I}}{2\pi}~.
\end{eqnarray}
Anticipating spin-charge separation (decoupling) in 1D, we define the independent charge and spin fluctuation fields in each wire which commute with each other:
\be 
\varphi_{c}^{a} = \frac{1}{\sqrt{2}} \left[\varphi_{\uparrow}^{a} +  \varphi_{\downarrow}^{a}\right],~~
\varphi_{s}^{a} = \frac{1}{\sqrt{2} } \left[\varphi_{\uparrow}^{a} -  \varphi_{\downarrow}^{a}\right]~ .
\ee
Using the notation, $\nu = c/s$ to represent either charge or spin degrees of freedom, these can in turn be used to form the pair of canonically conjugate phase fields in each wire: 
\be 
\phi_{\nu} = \frac{1}{\sqrt{2}} \left[\varphi_{\nu}^{O} + \varphi_{\nu}^{ I} \right]~, ~~ 
\theta_{\nu} = \frac{1}{\sqrt{2}} \left[\varphi_{\nu}^{I} - \varphi_{\nu}^{O}\right]~,
\ee
which, in a non-interacting theory satisfy the commutation relation $[\phi_{\nu, i}(x), \theta_{\nu, i}(x')] = -i \pi \Theta(x-x')$. The Hamiltonian for each wire, including the effect of short ranged e-e interaction is now expressed in terms of these fields as 
\be \label{H}
H = \int_{0}^{L} dx \sum_{\nu=c,s} \frac{v_\nu}{4 \pi} \left[  g_\nu (\nabla \phi_\nu)^2 +\frac{1}{g_{\nu}} (\nabla \theta_\nu)^2  \right]~, 
\ee
where  $v_c$ and $v_s$ denote the charge and spin velocities and $g_c$ and $g_s$ are the two interaction parameters for charge and spin sector, respectively, [\onlinecite{Giamarchi}] which we assume to be the same in all the wires. Note that 
$g_{c/s} = 1$ corresponds to the noninteracting case, $0 < g_c < 1$
to repulsive e-e interactions, and $g_s \neq 1$ to a broken
spin SU(2) symmetry. 
In the absence of an external magnetic field, or spin-dependent interactions, $g_s =1$ due to the underlaying SU(2) symmetry in the spin space. In Eq.~\eqref{H},   $L$ denotes the length of the wires, which is  assumed to be much longer than the width of the low-energy wave packets, $L k_F \gg 1$, which allows us to use a continuum description. Furthermore we have completely ignored phonons and disorder in the TLL wires.

The Hamiltonian in Eq.~\eqref{H}, which includes e-e interactions, can be diagonalized in terms of free bosonic fields in each wire, which satisfy the equal time canonical commutation relation: $[\tilde{\phi}_{\sigma}^{a}(x),\tilde{\phi}_{\sigma'}^{a'}(x')] = a i\pi \Theta(x-x') \delta_{\sigma \sigma'} \delta_{aa'}$. 
This is achieved by scaling the interacting fields in Eq.~\eqref{H} as
\begin{equation}\label{11}
{\phi}_{\nu}=\tilde\phi_{\nu}/\sqrt{g_{\nu}}~,~~~{\rm and }~~~ \theta_{\nu} =  \sqrt{g_{\nu}}~\tilde\theta_\nu~.
\end{equation}
The incoming and outgoing interacting fields in each wire,  can now be expressed in terms of the scaled free fields through the usual Bogoliubov transformation, given by 
\ber\label{12}
\varphi_{\sigma}^{a}& = &\frac{1}{2\sqrt{2 g_c}}\left[(1+g_c)\tilde{\varphi}_{c}^{a}+(1-g_c)\tilde{\varphi}_{c}^{\bar{a}}\right] \nonumber \\
& + & \sigma\frac{1}{2\sqrt{2 g_s}}\left[ (1+g_s)\tilde{\varphi}_{s}^{a}+(1-g_s)\tilde{\varphi}_{i}^{\bar{a}} \right]~.
\eer
In the above equation $\bar{a}$ is defined such that if $a=I$ then $\bar{a}=O$ and vice versa and $\sigma = \pm 1$ for up-spin and down-spin electrons respectively (when not used as a subscript).

\subsection{Bosonization of the junction}
In addition to the Hamiltonian in Eq.~\eqref{H} which describes each of the three disconnected spin-1/2 interacting wires,  to describe the junction we need to impose additional boundary conditions on the fields in each wire at the junction. As a consequence of the spin-charge separation in a spin-1/2 TLL wire, the charge and the spin sectors satisfy independent boundary conditions. 
Following the standard procedure [\onlinecite{chamon1, das2, agarwal_tdos,Hou}], 
we consider a current splitting matrix $\M$, which relates the incoming charge and spin currents to  the outgoing spin and charge currents (in the noninteracting theory), and consequently the incoming and outgoing bosonic fields, i.e, $j_{\nu,i}^{O}=\sum_{j}(\mathbb{M}_{\nu})_{ij}~j^{I}_{\nu,j}~$, which leads to $\phi_{\nu,i}^{O}=\sum_{j}(\mathbb{M}_{\nu})_{ij}~\phi^{I}_{\nu,j}~$ (where $\nu= c/s$ denotes the charge or the spin sector). Here we have neglected an integration constant which plays no physical role in the computation of TDOS exponent or in the scaling dimensions of various operators.  To enforce the condition that the boundary condition specified by the matrix $\M$ represents a fixed point of the theory, the incoming and outgoing fields must satisfy the usual bosonic commutation relations. This restricts the matrix $\M$ to be orthogonal and the orthogonality condition of $\M$ also implies that there is no dissipation at the junction [\onlinecite{agarwal_diss}]. Furthermore, the conservation of charge and spin current at the junction, ensure that each of the rows of the matrix $\M$ add up to unity.

For a junction of two or more TLL spin-1/2 wires, one has to impose independent boundary conditions on the charge and spin sectors, and thus all fixed points will be represented by a product of two matrices:  $\M_c \M_s$, with the first matrix specifying the boundary for the charge sector, and the second matrix specifying the boundary condition for the spin sector. 
For the case of a two-wire junction, there are only two possibilities for the $\M$ matrices which are given by,
\begin{equation}\label{14}
R_N = \left(\begin{matrix}
1 & 0 \\ 
0 & 1
\end{matrix} \right) 
\mbox{ and }
R_D = 
\left(\begin{matrix}
0 & 1 \\ 
1 & 0
\end{matrix} \right), 
\end{equation}
with $R_N$ representing the disconnected or reflecting fixed point (the subscript $N$ denotes Neumann boundary condition) and  $R_D$ representing the connected or transmitting fixed point respectively (the subscript $D$ stands for Dirichlet boundary condition), with both of them preserving time reversal symmetry (TRS). Now, all possible fixed point or boundary conditions for a two wire junction of spin-1/2 electrons are given by $R_{N} R_{N}$ (both charge and spin sectors disconnected), $R_{N} R_{D}$ (charge sector disconnected and spin sector connected), $R_{D} R_{D}$ (both charge and spin sectors connected) and finally $R_{D} R_{N}$ (charge sector connected and spin sector disconnected).

For a three wire charge and spin conserving Y junction, all current splitting orthogonal matrices $\M$ can be parametrized by a single continuous parameter $\theta$ as in the case of spin-less electron wires [\onlinecite{das2, agarwal_tdos}]. For spin-1/2 wires, there are two such continuous parameters $\theta_c$ and $\theta_s$, specifying the boundary condition (or fixed point) corresponding to the charge and spin degree of freedom respectively. The $3\times 3$ matrices $\M$ specifying the boundary condition for each sector, preserve time reversal symmetry (TRS), only if they are symmetric, and based on this they can be divided into two classes: det($\M_{1\nu})=1$ and det($\M_{2\nu})=-1$. These two classes have a form explicitly given by
\begin{equation}\label{15}
\M_{1\nu}=
\left(\begin{matrix}
a & b & c \\ 
c & a & b \\
b & c & a
\end{matrix} \right)_{\nu} , ~~
\M_{2\nu}=
\left(\begin{matrix}
b & a & c \\ 
a & c & b \\
c & b & a
\end{matrix} \right)_{\nu},
\end{equation}
where $a=(1+2\cos\theta_{\nu})/3, ~b=(1-\cos\theta_{\nu}+\sqrt{3}\sin\theta_{\nu})/3$, and $c=(1-\cos\theta_{\nu}-\sqrt{3}\sin\theta_{\nu})/3$. This implies four families of fixed points for a Y-junction of interacting spin-1/2 wires: $\M_{1c} \M_{1s}$, $\M_{1c} \M_{2s}$, $\M_{2c} \M_{1s}$ and finally $\M_{2c} \M_{2s}$. Of these only the $\M_{1c} \M_{1s}$ corresponds to a $Z_3$ symmetric (in the wire  indices) class of fixed points, while the other three specify an asymmetric class of fixed points (with broken $Z_3$ symmetry in the wire indices). 

The matrix $\M$ connects the incoming and outgoing field in a non-interacting theory. In presence of e-e interactions in the wire,  it  also has to undergo a Bogoliubov transformation:  $\M \to \widetilde{\M}$ so that it connects the incoming and outgoing effective `free' fields at the junction via the relation, $\tilde{\phi}_{\nu,i}^{O}(0,t)=\sum_{j}(\widetilde{\M}_{ \nu})_{ij}~\tilde{\phi}_{\nu,j}^{I}(0,t)$. The Bogoliubov transformed matrix is  given by [\onlinecite{das2, agarwal_tdos}],
\begin{equation}\label{13}
\widetilde{\M}_{\nu}=[(1+g_\nu)\mathbb{I}-(1-g_\nu)\M_{\nu}]^{-1}[(1+g_\nu)\M_\nu-(1-g_\nu)\mathbb{I}]~.
\end{equation}
The $\M_{2\nu}$ class of matrices have an interesting property: $(\M_{2\nu})^2 = \mathbb{I}$. As a consequence $\widetilde{\M}_{2\nu}=\M_{2\nu}$, which implies that at the junction both interacting and free fields have identical properties. For $\M_{1\nu}$, the matrix $\widetilde{\M}_{1\nu}$ still has the same form as $\M_1$, but with the corresponding matrix elements given by $\tilde{a}=(3g^2_{\nu}-1+(3g^2_{\nu}+1)\cos\theta_{\nu})/\delta$ and $\tilde{b}/\tilde{c}=2(1-\cos\theta_{\nu}\pm \sqrt{3}g_{\nu}\sin\theta_{\nu})/\delta$, where $\delta=3[1+g_{\nu}^2+(g^2_{\nu}-1)\cos\theta_{\nu}]$.  Note that the formulation of delayed evaluation of boundary condition described in this section and Eq.~\eqref{13} is applicable to any number of interacting one-dimensional wires connected at a  dissipation-less junction described by a bosonic fixed point.

A detailed and systematic study of the stability of various fixed points for a Y-junction of spin-1/2 electrons has been done in Ref.~[\onlinecite{Hou}], using conformal field theory as well as bosonization with delayed evaluation of boundary conditions,   in the $g_s-g_c$ parameter space and several important fixed points with a unique attractive basin have been identified. For the sake of completeness, and as a check of our calculations, we report the scaling dimensions of all the spin-preserving single particle and pair tunneling operators, for all possible fixed points in Appendix. 

For the $\M_{1c} \M_{1s}$ class of fixed points, $\theta_{c/s}=\pi$ or $[a,b,c]=[-1/3,2/3,2/3]$, represents the so called $DD$ (Dirichlet-Dirichlet) fixed point for charge/spin sector, $\theta_{c/s}=0$ or $[a,b,c]=[1,0,0]$ indicates the $NN$ (Neumann-Neumann) or disconnected fixed point in which there is no tunneling between any pair of wires. The case of $\theta_{c/s}= \frac{2\pi}{3}$ ($[a,b,c]=[0,1,0]$) and $\theta_{c/s}= \frac{4\pi}{3}$ ($[a,b,c]=[0,0,1]$)  corresponds to the chiral $\chi^{-}\chi^{-}$ and $\chi^{+}\chi^{+}$ fixed points, respectively. Other important fixed points belonging to the $\M_{1c} \M_{1s}$ class with a unique attractive basin are specified as DN for $\theta_c = \pi$ and $\theta_s = 0$ and ND for $\theta_c = 0$ and $\theta_s = \pi$.

For the $\M_{2c} \M_{2s}$ class of fixed points, $\theta_{c/s}=0,$  $2 \pi /3$ or $4 \pi/3$  
represents the asymmetric set of fixed points called $D_AD_A$, where two of the three wires are connected in both the charge and spin sectors while one of the wires is completely disconnected. All the three cases are identical and we will discuss the case of $\theta_c = \theta_s = 0$, where wires 1 and 2 are connected in both the charge and spin sectors and wire 3 is completely disconnected.  Note that  even though the $D_A D_A$ fixed point is asymmetric in the wire indices, it preserves TRS since only two of the wires are connected. There are several other interesting fixed points as well, however only the ones discussed here have a unique basin of attraction in the $g_s-g_c$ plane.  Along the SU(2) invariant line, $g_s =1$, the NN fixed point is stable for $g_c <1$, the chiral fixed point is stable for $1<g_c <3$, and the DN stable fixed point is stable for $g_c >3$ [\onlinecite{Hou}]. 

Finally, we note that the matrices $\M_\nu$ specifying the boundary condition for the charge and spin degrees of freedom at the junction are also associated with the charge and spin conductance tensors associated with each fixed point. The outgoing charge and spin current in wire $j$ is given by $I_j^\nu = G_{jk}^{\nu} V_k^\nu$, 
where the superscript  $\nu = c/s$, and $V^c_k$ is the voltage applied on wire $k$, and $V^s_k = \mu_{k \uparrow} - \mu_{k \downarrow}$ is the chemical potential difference between the up and down spin electrons in wire $k$.  If the TLL wires are connected to Fermi-liquid leads, then the charge and spin conductance tensor is 
\begin{equation} \label{eq:GFL}
\mathbb{G}^\nu_{\rm FLL}=\frac{2 e^2}{h}\left(\mathbb{I}-\M_\nu\right)~. 
\end{equation}
If there are no Fermi liquid leads, and the TLL wires extend to infinity, then the charge and spin conductance tensor is given by 
\begin{equation} \label{eq:GTL}
\mathbb{G}^\nu_{\rm TLL}=\frac{2 g_\nu e^2}{h}\left(\mathbb{I}-\widetilde{\M}_\nu\right)~. 
\end{equation}
We note here that Eq.~(\ref{eq:GFL}) and Eq.~(\ref{eq:GTL}), are strictly valid in equilibrium and at zero temperature. Any deviation from the equilibrium (say arising from a finite bias voltage or finite temperature) will lead to $g$ dependent power law corrections, either on temperature or on the bias voltage or the system size, depending on whichever corresponds to the shortest 
length scale in the problem as in the case of a two wire junction [\onlinecite{Safi3}]. 

Having described the various fixed points at the junction and the associated conductance for both the Fermi liquid and TLL leads, we now proceed to calculate the TDOS of the spin-1/2 Y-junction.
\section{Tunneling density of states}
\label{secIII}
In this section we compute the TDOS of a spin-1/2 Y-junction for adding an electron with energy $\hbar\omega$ in one of the wires. The expression of the total TDOS in $i^{\rm th}$ wire is given by the sum of the spin resolved TDOS  $\rho_i(\omega) = \rho_{\uparrow i} (\omega) + \rho_{\downarrow i} (\omega)$, where 
\begin{equation}\label{16}
\rho_{\sigma i}(\omega) \equiv \frac{1}{2 \pi} \int_{-\infty}^{\infty} e^{i\omega t}  \left\langle\{\psi_{\sigma i}(x,t), \psi_{\sigma i}^{\dagger}(x,0)\}\right\rangle dt ~.
\end{equation}
The spin resolved Green's function in the $i^{\rm th}$ wire is $G=\langle\psi_{\sigma i}(x,t)\psi_{\sigma i}^{\dagger}(x,0)\rangle$. For a system with long wires, i.e., in the  $ L\rightarrow \infty$ the Green's function can be written as $G=G_I+G_O=\langle\psi^{I}_{\sigma i}(x,t)\psi_{\sigma i}^{I\dagger}(x,0)\rangle+\langle\psi^{O}_{\sigma i}(x,t)\psi_{\sigma i}^{O\dagger}(x,0)\rangle$, 
where we have neglected two  oscillatory terms which vanish in the $L \to \infty$ limit,  and are unimportant.
The non oscillatory terms in the spin resolved Green's function are explicitly given by 
\begin{eqnarray}\label{17}
& &\langle\psi^{O}_{\sigma i}(x,t)\psi_{\sigma i}^{O\dagger}(x,0)\rangle=
\langle\psi^{I}_{\sigma i}(x,t)\psi_{\sigma i}^{I\dagger}(x,0) \rangle    \\
& & =\frac{1}{2 \pi \alpha} \prod_{\nu = c, s} \left[\frac{-i \alpha}{v_\nu t - i \alpha} \right]^{\beta_\nu}
\left[\frac{-\alpha^2-4x^2}{(v_\nu t- i\alpha)^2-4x^2}\right]^{\gamma_\nu \tilde{d}_{\nu i}}~, \nonumber
\end{eqnarray}
where we have defined $\beta_\nu \equiv (1+g_\nu^{2})/(4g_\nu)$, $\gamma_\nu \equiv (1-g_\nu^2)/(8 g_\nu)$ and  $\tilde{d}_{\nu i}$ denotes the $i^{\rm th}$ diagonal elements of the corresponding $\widetilde{\mathbb{M}}_{\nu}$ matrix representing the boundary condition at the junction and $\nu = c/s$.  
For fixed points belonging to the $\M_{1c}\M_{1s}$ class of fixed points $\tilde{d}_{\nu i}=\tilde{a}_{\nu}$, and for fixed points belonging to the $\M_{2c} \M_{2s}$ class of fixed points, $\tilde{d}_{\nu i}= \tilde{b}_{\nu},~\tilde{c}_{\nu},~\tilde{a}_\nu$ for wires one, two, and three, respectively.   Note that far away from the junction, in the $x \to \infty$ limit, the last two terms in Eq.~\eqref{17} become unity and do not contribute to the TDOS, while in the $x \to 0$ limit, they contribute to the TDOS making the TDOS exponent different at the junction [\onlinecite{Eggert}].

Inserting Eq.~\eqref{17} in Eq.~\eqref{16} and performing the integration we obtain analytical expressions for the TDOS for the limiting cases of  
$x\rightarrow0$ (in vicinity of the junction) and   $x\rightarrow\infty$ (far away from the junction). In both cases,  the TDOS for $\omega >0$ has the same analytical form and is given by 
\begin{equation}\label{18}
\rho_{\sigma i}(\omega)=\frac{1}{\pi \hbar \alpha \Gamma (\Delta_{i})}\tau_c^{\Delta_{i}}\omega^{\Delta_{i}-1} e^{-\omega \tau_c}~,
\end{equation}
where $\Gamma(x)$ represents the gamma function, $\tau_c=\alpha/v$ is an effective short time (inverse high frequency) cutoff, and  $\omega=eV/\hbar$, where $V $ is the bias voltage between the STM tip and the $i^{\rm th}$ wire [see Fig.~\ref{fig1}]. In Eq.~\eqref{18}, 
$\Delta_{i}$ is the spin independent TDOS exponent which can be expressed as a combination of two terms: $\Delta_i= \Delta_{ci} +  \Delta_{si}$, where $\Delta_{ci}$ ($\Delta_{si}$) is only  a function of $g_c$ ($g_s$) and $\theta_c$ ($\theta_s$) which specifies the boundary condition of the charge (spin) sector at the junction.

Note that the TDOS exponent in general depends on whether it is being measured far away from the junction, in the vicinity of the junction or at intermediate locations. Below we describe each of these regimes in subsections \ref{FA}, \ref{Near}, and \ref{IR}, respectively.
\subsection{TDOS exponent far away from the junction}
\label{FA}
Far away from the junction, i.e., in the $x\rightarrow\infty$ limit, for all possible classes of fixed points we find that 
\be \label{inf}
\Delta_{\infty} = \frac{1}{4}\left(g_c+\frac{1}{g_c}\right)+\frac{1}{4}\left(g_s+\frac{1}{g_s}\right)~,
\ee
independent of the boundary condition ($\theta_{c/s}$) at the junction, as expected. Equation~\eqref{inf}  also specifies the TDOS exponent for a single infinite wire made of spin-1/2 TLL and is well known in the literature [\onlinecite{Voit, Eggert,  Braunecker}].
We emphasize that  even  for the case of $g_s \neq 1$ where the SU2 symmetry is broken in the wires,  the TDOS exponent is identical for the up-spin and down-spin electron tunneling since the Hamiltonian for each wire in Eq.~\eqref{H}, and the boundary conditions at the junction are invariant under the interchange of spins, i.e., $\uparrow$-spin  $\to$ $\downarrow$-spin and vice versa.  Note that the result for the spineless case can be reproduced by substituting $g_c \to g$ and $g_s \to g$ in Eq.~\eqref{inf}.

\begin{figure}[t]
\centering
\includegraphics[width=.99\linewidth]{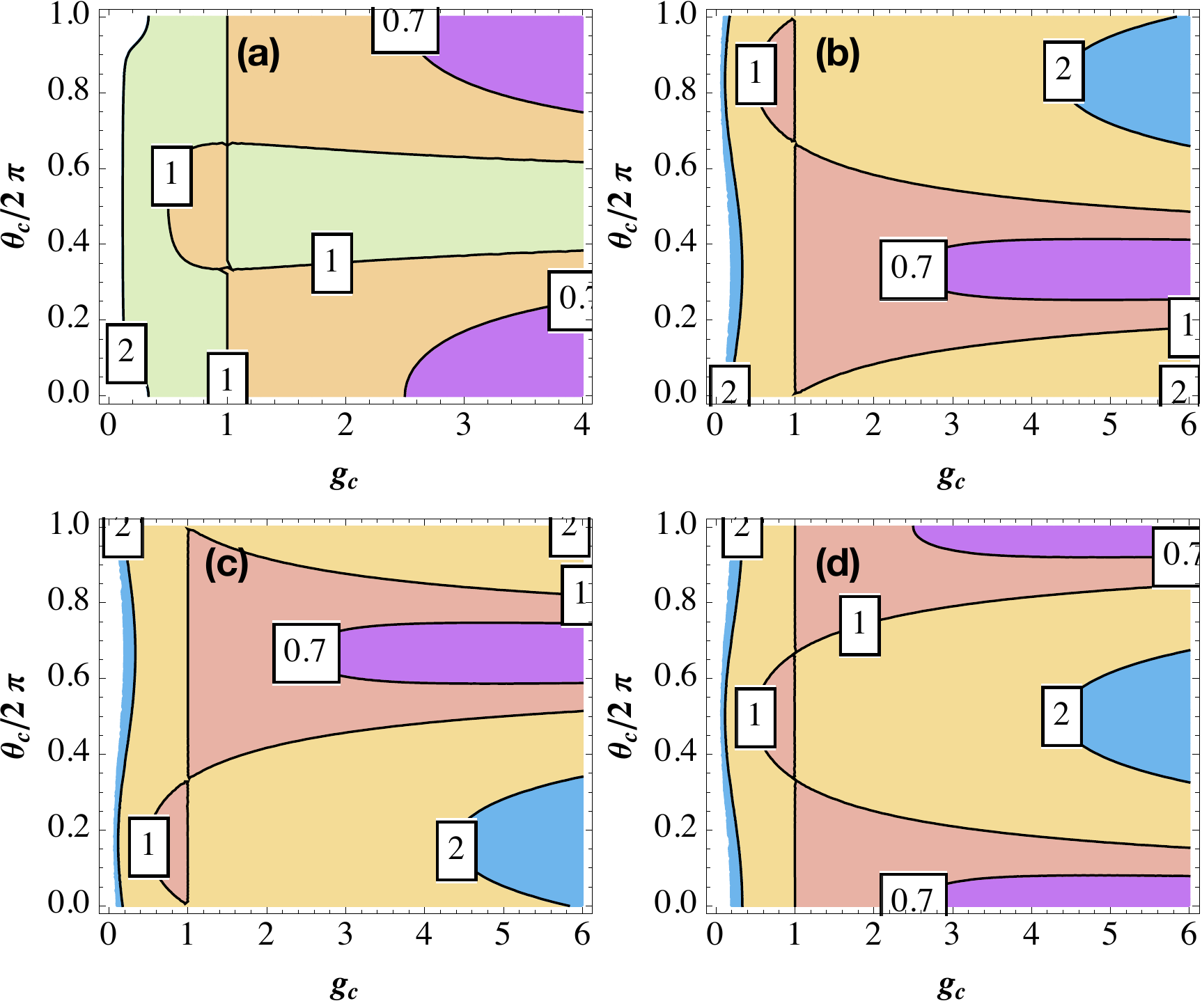}
\caption{Contour plot of the TDOS exponents in the vicinity of the junction, $\Delta_0$, $\Delta_1$, $\Delta_2$ and $\Delta_3$ in the $\theta_c-g_c$ plane for the SU(2) symmetric line $g_s=1$, in panels a), b), c), and d),
respectively. Note that for $g_s=1$, the TDOS exponents in Eqs.~\eqref{19} and~\eqref{20} are independent of $\theta_s$. The fixed points corresponding to the  dome shaped regions lying to the left of the $g_c=1$ vertical line in all four panels show an anomalous enhancement of the TDOS, even for repulsive interactions. The boxed numbers specify the numerical value of the exponent for the corresponding contour lines.
\label{fig2}}
\end{figure} 

\subsection{ TDOS exponent in the vicinity of the junction}
\label{Near}
In the vicinity of the junction,  $x\rightarrow0$,   $\Delta_{i}$ depends on the boundary conditions of both the charge ($\theta_c$) and the spin ($\theta_s$) sectors in general. For boundary conditions where both the charge and the spin sectors belong to the  $\M_{1c} \M_{1s}$ class, the Y-junction has Z$_3$ symmetry in the wire indices and the TDOS exponent is identical in all the wires. The TDOS exponent is given by $\Delta_0=\Delta_{0c} + \Delta_{0s}$, where  
\begin{eqnarray}\label{19}
\Delta_{0\nu}&=&\frac{1}{6g_{\nu}}\frac{1+5g_{\nu}^2+(g_{\nu}^2-1)\cos\theta_{\nu}}{1+g_{\nu}^2+(g_{\nu}^2-1)\cos\theta_{\nu}}~, 
\end{eqnarray}
where $\nu = c/s$.
As a consistency check we note that as $g_\nu \to g$, and $\theta_\nu \to \theta$,  the TDOS exponents for the spin-1/2 TLL Y-junction become identical to the case of spinless three wire junction, reported in Ref.~[\onlinecite{agarwal_tdos}]. 

For the case of broken Z$_3$ symmetry, i.e., where both or either of the charge and spin sector boundary conditions at the junction are specified by the $\M_2$ class of matrices,  the TDOS exponent, in the vicinity of the junction, explicitly depends on the wire index $i$.  Particularly for the 
$\M_{2c} \M_{2s}$ class of fixed points, the exponent is given by $\Delta_{i}=\Delta_{ic}+\Delta_{is}$, where the spin and charge part of the exponent for wire $1$ are explicitly given by 
\be \label{20}
\Delta_{1\nu}=\frac{4+2g_{\nu}^2+(\cos\theta_{\nu}-\sqrt{3}\sin\theta_{\nu})(g_{\nu}^2-1)}{12g_{\nu}}~,
\ee
For the other two wires, $\Delta_{2\nu}$ and $\Delta_{3\nu}$ are simply obtained by shifting $\theta_{\nu} \rightarrow \theta_{\nu} \mp 2 \pi /3$ respectively,  in the corresponding expressions for $\Delta_{1\nu}$ in Eq.~\eqref{20}. 
For the $\M_{1c}\M_{2s}$ class of fixed points, the TDOS exponent is given by $\Delta_{i}=\Delta_{0c}+\Delta_{is}$, and for the $\M_{2s}\M_{1c}$ class of fixed points, it is given by $\Delta_{i}=\Delta_{ic}+\Delta_{0s}$, where $\Delta_{0\nu}$ is given in Eq.~\eqref{19}, and $\Delta_{i \nu}$ is specified by Eq.~\eqref{20}. 

For the particular case of the SU(2) symmetric line $g_s =1$, the spin part of the TDOS exponent in both Eqs.~\eqref{19} and~\eqref{20} become	s $\Delta_{js}  =1/2$,  independent of $\theta_s$, where $j = 0, 1, 2{~\rm or}~3$. We plot the TDOS exponent for this particular case, in the $\theta_c -g_c$ plane in Fig.~\ref{fig2}. In all four panels, the dome shaped region to the left of the $g_c =1$  vertical line, indicates fixed points showing an anomalous enhancement of the TDOS for small bias voltage even for repulsive interactions, in the vicinity of the spin-1/2 Y-junction. 

To gain more insight into the behaviour of the fixed points corresponding to the TDOS enhancement in the weekly interacting region ($g_c \approx 1$) for the $g_s =1$ line, we expand the TDOS exponents to the lowest order in the small parameter $(1-g_c)$ and obtain, $\Delta_j = 1+ (1-g_c) d_{jc}$/2, where $d_{jc} = a_{c},b_{c}, c_{c}{~\rm or}~ a_c$ for $j = 0,1,2{~\rm or}~ 3$, and it simply denotes the diagonal elements of the corresponding $\M_c$ matrix,  which relates the non-interacting incoming and outgoing charge fields at the junction. Thus, for weekly repulsive interactions, $g_c <1$, we find that $\Delta_j$ is less than one (or TDOS enhancement occurs) wherever \textcolor{blue}{$d_{jc} <0$}, which physically corresponds to a hole current being reflected from the junction for an incident electron current. This is consistent with the previously reported enhancement of the TDOS for a TLL wire connected at one end to a superconductor [\onlinecite{Fazio}], where a proximity effect induced Andreev process leads to the reflection of a hole for an incident electron. We emphasize here that even though we do not have any superconductivity explicitly in our model,  the current splitting matrices, span all possible scenarios and includes the cases where a hole (either fully or partially) is reflected at the junction.

\begin{figure}[t]
\centering
\includegraphics[width=1.0\linewidth]{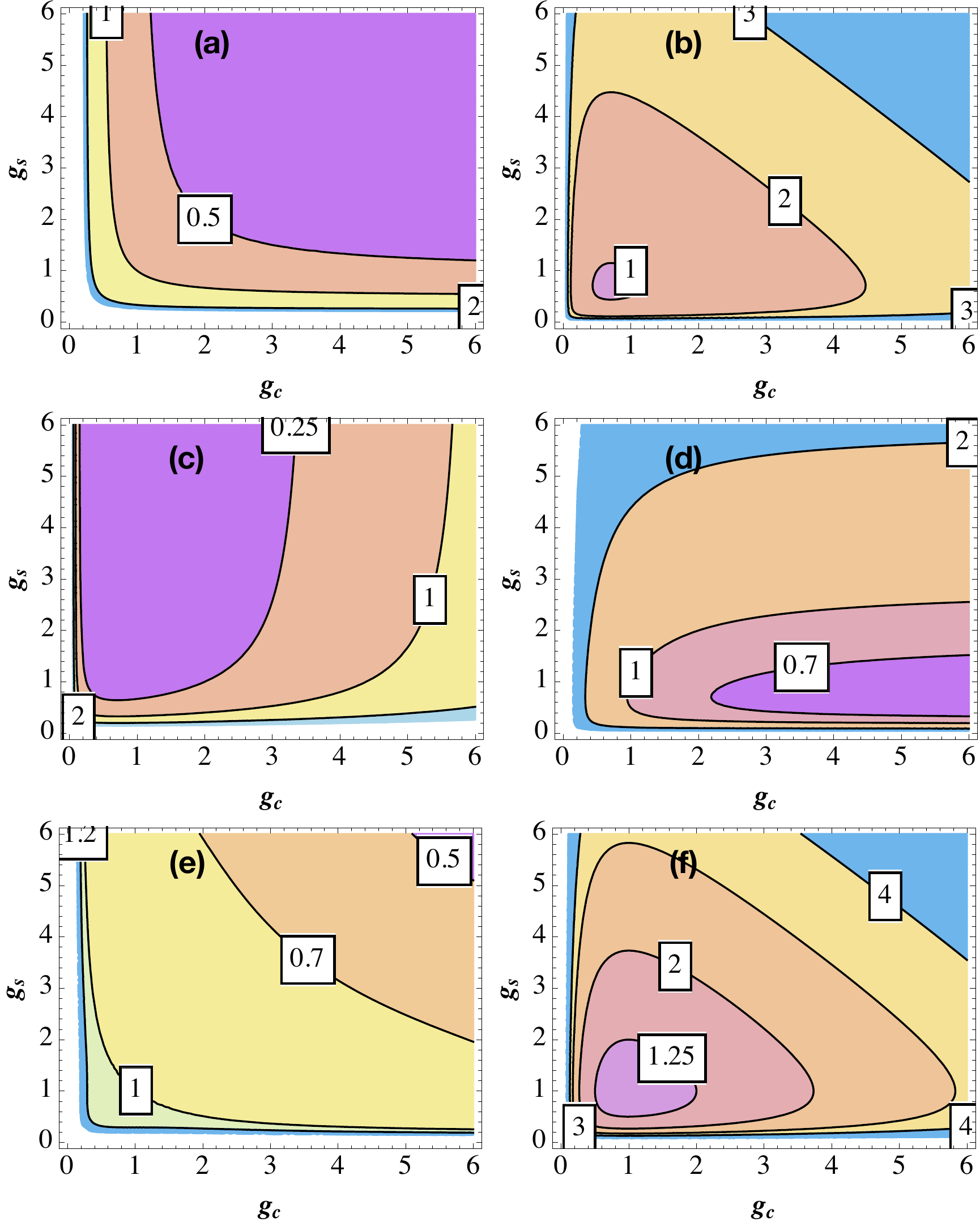}
\caption{Contour plot of the TDOS exponents in the $g_s-g_c$ plane for various fixed points. Panels a), b), c), d), and e) display $\Delta_0$ for  various fixed points of $\M_{1c} \M_{1s}$ type, i.e., 
NN ($\theta_c=\theta_s=0$), DD ($\theta_c=\theta_s=\pi$), DN ($\theta_c=\pi, \theta_s=0$), ND ($\theta_c=0, \theta_s=\pi$), and $\chi^+\chi^+$ ($\theta_c=\theta_s=2\pi/3$) respectively. Panel f) denotes TDOS exponent $\Delta_1$ for the D$_A$D$_A$ fixed point of the $\M_{2c} \M_{2s}$ type. 
The region bounded by $\Delta_i <1$, in all the panels signifies TDOS enhancement in the zero bias limit and we clearly see enhancement in the TDOS even for repulsive e-e interactions ($g_c <1$) in several cases. The boxed numbers specify the numerical value of the exponent for the corresponding contour lines.
\label{fig3}}
\end{figure} 

Let us now consider some specific fixed points for the three wire junction of spin-1/2 electrons, which have a unique basin of stability in the $g_c-g_s$ plane [\onlinecite{Hou}], starting with the  fixed points belonging to the $\M_{1c}\M_{1s}$ class. For the NN fixed point 
$\Delta_0 = 1/(2 g_c) + 1/(2 g_s)$, and for $g_s=1$ the range of $g_c$  where $\Delta_0 <1$, or  TDOS enhancement occurs, is given by $g_c > 1$. Since the NN fixed point represents three disconnected wires, this result is consistent with earlier results for both spin-1/2  [\onlinecite{Eggert}] and the spinless disconnected wires [\onlinecite{delft, Furusaki, Fabrizio, Oreg}] where the TDOS enhancement at the ends of a single TLL wire can only occur  for attractive interactions ($g_c~ {\rm or}~ g >1$). For the DD fixed point ($\theta_\nu = \pi$), $\Delta_0 = (1+2g_c^2)/(6g_c) + (1+2g_s^2)/(6g_s) $, which implies that for $g_s =1$ the TDOS enhancement occurs for $1/2 < g_c < 1$, i.e., even for repulsive interactions.  For the DN fixed point ($\theta_c = \pi$ and $\theta_s = 0$)  $\Delta_0 = (1+2g_c^2)/(6g_c) +1/(2 g_s)$, and like the DD fixed point, the TDOS enhancement for the SU(2) symmetric line $g_s =1$ occurs for  $1/2 < g_c < 1$. In case of the ND fixed point ($\theta_c = 0$ and $\theta_s = \pi$), the TDOS exponent is $\Delta_0 = 1/(2 g_s) + (1+2g_s^2)/(6g_s)$, and for $g_s=1$ TDOS enhancement only occurs for attractive interactions as in the NN case, i.e. for $g_c > 1$. 
For the chiral fixed points, $\chi^+ \chi^+$ and $\chi^- \chi^-$, $\Delta_0 =  (1+3 g_c^2)/(2 g_c^3 + 6 g_c) + (1+3 g_s^2)/(2 g_s^3 + 6 g_s)$, and the TDOS enhancement again occurs only for $g_c >1$ for $g_s =1$. 

Another fixed point with a unique basin of stability is the  D$_A$D$_A$ fixed point, which belongs to the $\M_{2c} \M_{2s}$ class and is Z$_3$ asymmetric in the wire indices. Let us specifically consider the case of $\theta_c = \theta_s = 0$, in which wires 1 and 2 are fully connected in both the charge and spin sectors and wire 3 is completely disconnected. For this case $\Delta_1 = \Delta_2= \Delta_\infty = (1+g_c^2)/(4 g_c) + (1+g_s^2)/(4 g_s)$, as expected in the bulk of an infinite TLL wire,  and $\Delta_3 = 1/(2 g_c) + 1/(2 g_s)$ as expected, consistent with TDOS exponent near the ends of a TLL wire in Ref.~[\onlinecite{Eggert, Blumenstein}]. However, in this case, there is no anomalous TDOS enhancement  either in wires 1 and 2 for any value of $g_c$, or in wire 3 for repulsive interactions, along the $g_s=1~$ line.  
In Fig.~\ref{fig3}, we plot the spin independent TDOS exponent in wire 1  ($\Delta_{1}$) in the $g_s - g_c$ plane for several fixed points (NN, DD, DN, ND, $\chi^+ \chi^+$ and D$_A$D$_A$) and clearly see regions where the TDOS is enhanced ($\Delta_1 < 1 $) in the zero bias limit, even for repulsive e-e interactions. 

Note that the D$_A$D$_A$ fixed point considered above also describes the two wire junction $R_D R_D$ where both charge and spin sectors are connected (wires 1 and 2), and the $R_N R_N$ case where both the charge and spin sectors are disconnected (wire 3). The case of $R_D R_N$ fixed point where the charge sector is connected and the spin sector is disconnected can be constructed by considering wires 1 and 2 in the $\M_{2c}\M_{1s}$ class with $\theta_c =\theta_s = 0$ (also called the D$_A$N fixed point in the three wire context) and the TDOS exponent in this case is given by $\Delta_1 = \Delta_2 = (1+g_c^2)/(4 g_c) + 1/(2 g_s)$, which for $g_s \le 1$ can never have TDOS enhancement for any value of $g_c$.  Finally, the case of $R_N R_D$ fixed point with the spin sector connected while the charge sector is disconnected, is equivalent to considering wires 1 and 2 in the ND$_A$ fixed point (of $\M_{1c} \M_{2s}$ class) with $\theta_c =\theta_s = 0$, and in this case $\Delta_1 = \Delta_2 = 1/(2 g_c) + (1+g_s^2)/(4 g_s) $, which for $g_s =1$ can have TDOS enhancement only with attractive interactions ($g_c >1$).

\subsection{TDOS crossover from boundary to bulk at finite distance from the junction}
\label{IR}
At finite distance from the junction, $x \neq 0$,  TDOS can be obtained from Eq.~\eqref{16} after substituting Eq.~\eqref{17}, and performing the integration over $t$ in the upper half complex plane, along the five branch cuts running parallel to the imaginary axis with branching points $ i\alpha$, $i \alpha \pm 2x/v_c $, and $i \alpha \pm 2x/v_s $. A similar approach is described in Appendix A of Ref.~[\onlinecite{Liu}] where the authors studied TDOS in a spiral TLL wire in proximity to a superconductor.  For $ 0 < \omega<\hbar v_{\rm min}/\alpha $, where $v_{\rm min} = {\rm min}\{v_c,v_s\}$, only the first term in Eq.~\eqref{16} contributes, 
and the TDOS integral in Eq.~\eqref{16} becomes equivalent to the sum of five contour integrals around  each of the branch cuts.  The TDOS asymptote at finite $x > v_{\rm min}/(2 \omega)$ is obtained to be $\rho_{i \sigma}(x, \omega) \approx \rho_\infty  + \rho_{i\sigma}^{(c)} +  \rho_{i\sigma}^{(s)}$, where $\rho_\infty$ does not depend on the spatial coordinate and it is given by Eqs.~\eqref{18} and~\eqref{inf}. For $x>v_{\rm min}/(2\omega)$,  the other two terms display an oscillatory power law behaviour on the spatial coordinate and are explicitly given by 
\ber \label{xdep}
\rho_{i \sigma}^{(\nu)} (x, \omega)   & = & \frac{ 2^{1-\Delta_{\infty}}~ v_\nu^{\beta_{\bar{\nu}} + \gamma_{\bar{\nu}} \tilde{d}_{\bar{\nu} i} }}{\pi v_\nu \Gamma(\gamma_\nu \tilde{d}_{\nu i})  v_{\bar{\nu}}^{\beta_{\bar{\nu}}} (v_\nu-v_{\bar{\nu}})^{\gamma_{\bar{\nu}} \tilde{d}_{\bar{\nu} i} } }   
 \left(\frac{\omega \alpha}{v_\nu}\right)^{\gamma_{\nu} \tilde{d}_{\nu  i} } \nonumber \\
& \times & \left(\frac{x}{\alpha}\right)^{ \gamma_\nu \tilde{d}_{\nu  i} - \Delta_\infty }  \cos\left(\frac{2x \omega}{v_\nu} - \delta_\nu\right)~,
\eer 
where $\delta_\nu = {\rm{Arg}}(i^{\Delta_\infty + \gamma_\nu \tilde{d}_{\nu  i}})$,  $(\nu, {\bar \nu}) = (c,s)~ {\rm or}~ (s,c)$, and $\beta_\nu$, $\gamma_\nu$ and $\tilde{d}_{i \nu}$ are defined below Eq.~\eqref{17}. Along the SU(2) symmetric line, $g_s =1$, $\gamma_{s} =0$ and the $\gamma_s$ dependent term in Eq.~\eqref{18} drops out of the TDOS integral, and the spatial dependence of the TDOS arises only from the charge term: $\rho_i^{(c)} \propto \cos(2x \omega/v_c - \delta_c) x^{\gamma_c \tilde{d}_{ci} - \beta_c -1/2}$ with a slowly oscillating contribution that drops off as a power law. Note that in the vicinity of the junction [$x \ll  v_{\rm min}/(2 \omega)$], the TDOS has a power law behaviour: $\rho_i \propto \omega^{\Delta_i -1}$ where $\Delta_i$ depend on the boundary conditions at the junction [see Eqs.~\eqref{19} and \eqref{20}] . For small distances away from the junction [$x \approx  v_{\rm min}/(2 \omega)$], TDOS  shows an oscillatory behaviour with bias frequency and for large distances away from the junction [$x \gg  v_{\rm min}/(2 \omega)$], the power law behaviour in $x$ [see Eq.~\eqref{xdep}] suppresses the oscillations and the TDOS goes over to the bulk value, $\rho_i \propto \omega^{\Delta_{\infty} -1}$. 

\section{Conclusion}
\label{summary} 
To summarize, in this article we explicitly calculate the local tunneling density of states in the vicinity of a spin-1/2 TLL Y-junction, and present an analytic expression for the TDOS exponent in terms of the boundary condition at the junction and the strength of e-e interaction. We find that there are several fixed points which in the vicinity of the junction give an anomalous TDOS enhancement in the zero bias limit, even for repulsive interactions. Physically all such instances of TDOS enhancing fixed points are such that  holes are reflected from the junction for  incident electrons. This makes the TLL junction physically similar to the case of a TLL connected to a superconductor where the TDOS enhancement was attributed to the proximity induced Andreev processes [\onlinecite{Fazio}], even though the three wire junction considered by us has no superconductor. 
 
 It should be noted that the TDOS expression in the vicinity of the junction, given by  Eq.~\eqref{18},  is valid only for $x< x_c = v_{\rm min}/(2\omega)$, where $v_{\rm min} = {\rm min}\{v_c, v_\sigma \}$. For a realistic system, $v_{\rm min} \approx 10^5$ m/s,  and for a STM tip voltage of $1$mV, the crossover length scale is $x_c \approx 33$ nm, which is within current experimental reach. For $x \approx x_c$, the TDOS displays an oscillatory behaviour which is again suppressed at large distances, $x \gg x_c$, and the TDOS reverts back to its bulk value. One limitation of our calculations is that they are valid only for the  regime of bias frequencies which do not breach the linearization regime of each TLL wire, i.e., $\omega <  v_{\rm min}/\alpha$.
 
Note that while we have considered a junction with symmetric e-e interaction strength, in a more realistic situation this may not be the case due to asymmetrical screening induced by gates, inhomogeneities, or defects. 
A detailed analysis of TDOS in a Y-junction with different values of $g$ in the three wires, can be done in the spirit of Ref.~[\onlinecite{Hou_prb2012}], which focused on the stability and analysis of the fixed points.  However it is beyond the scope of the present paper, and can be the subject of a future work.  Additionally, we note that all our results are valid only in the regime where backward and umklapp interactions can be safely ignored [\onlinecite{CNTLL}]. 

In addition to the spin degree of freedom, our TDOS exponent calculations can also be extended to include other degrees of freedom such as different valleys in  carbon nanotubes. If the spin and the valley rotation symmetry are not broken, then all the corresponding TDOS exponents for carbon nanotube Y-junctions  are easily calculated using the fact that three of the four bosonic fields get pinned and only the center of mass field which corresponds to the charge degree of freedom primarily contributes interaction dependent term in the TDOS. The TDOS exponents for carbon nanotube Y-junctions are explicitly given by 
$\Delta_{j}^{\rm CN} = (2\Delta_{jc} + 3)/4$,  where $j = {0,1,2~\rm or ~3}$, and $\Delta_{jc}$ are specified by Eqs.~\eqref{19}-\eqref{20}.  This is consistent with the TDOS exponents reported for the bulk and the edge of a TLL hosted in a carbon nanotube [\onlinecite{CNTLL}].

Experimentally, spin-1/2 TLL wire junctions can be realized by means of carefully patterned 1D wires in a 2DEG, or via crossed single-walled nanotubes [\onlinecite{fuhrer}] and tuned to various fixed points by means of nano-gates applied near the junction. Another feasible possibility is an island like set-up of quantum Hall edge states, proposed in Ref.~[\onlinecite{das2}], in which the tunneling operators can be controlled via  gate voltage operated constrictions in the central region of the island. Once the junction is tuned to an appropriate fixed point, the power law exponent of the TDOS can be experimentally extracted by measuring the differential tunneling conductance as a function of the STM tip voltage (for fixed temperature), or as a function of temperature (for fixed voltage) as was done in Ref.~[\onlinecite{Blumenstein}]. 
\vspace{0.5cm}

\section*{Acknowledgments} We thank Diptiman Sen and Sumathi Rao for stimulating discussions and for carefully reading the manuscript. 
We gratefully acknowledge funding 
from the INSPIRE faculty fellowship by DST (Govt. of India), and from the 
Faculty Initiation Grant by IIT Kanpur, India. 

\begin{table}[t]
\centering
\caption{Scaling dimension of various single particle tunneling operators: $\delta_{0\sigma} = \delta_{kc} +  \delta_{ls}$, where $k,l = 1 {~\rm or~} 2$ and the boundary condition at the junction is specified by $\M_{kc} \M_{ls}$.
\vspace{0.5cm}
\label{T1}}
\begin{tabular}{c  c}\toprule
\vspace*{2mm}
Operator   & $\delta_{1\nu}  $ \\
\hline
\vspace*{2mm}
$\psi^{O~\dagger}_{i,\sigma}\psi^{I}_{i,\sigma}$   & $\frac{2g_{\nu}(1-\cos\theta_{\nu})}{3[g_{\nu}^2+\cos\theta_{\nu}(g_{\nu}^2-1)+1]}$ \\

\vspace*{2mm}
$\psi^{O~\dagger}_{2,\sigma}\psi^{I}_{1,\sigma},\psi^{O~\dagger}_{3,\sigma}\psi^{I}_{2,\sigma},\psi^{O~\dagger}_{1,\sigma}\psi^{I}_{3,\sigma}$ & $\frac{g_{\nu}(\cos\theta_{\nu}+\sqrt{3}\sin\theta_{\nu}+2)}{3[g_{\nu}^2+\cos\theta_{\nu}(g_{\nu}^2-1)+1]} $\\

\vspace*{2mm}
$\psi^{O~\dagger}_{1,\sigma}\psi^{I}_{2,\sigma},\psi^{O~\dagger}_{2,\sigma}\psi^{I}_{3,\sigma},\psi^{O~\dagger}_{3,\sigma}\psi^{I}_{1,\sigma}$   & $\frac{g_{\nu}(\cos\theta_{\nu}-\sqrt{3}\sin\theta_{\nu}+2)}{3[g_{\nu}^2+\cos\theta_{\nu}(g_{\nu}^2-1)+1]} $\\

\vspace*{2mm}
$\psi^{I~\dagger}_{2,\sigma}\psi^{I}_{1,\sigma},\psi^{I~\dagger}_{3,\sigma}\psi^{I}_{2,\sigma},\psi^{I~\dagger}_{1,\sigma}\psi^{I}_{3,\sigma} $& $\frac{g_{\nu}}{g_{\nu}^2+\cos\theta_{\nu}(g_{\nu}^2-1)+1} $\\

\vspace*{2mm}
$\psi^{O~\dagger}_{2,\sigma}\psi^{O}_{1,\sigma},\psi^{O~\dagger}_{3,\sigma}\psi^{O}_{2,\sigma},\psi^{O~\dagger}_{1,\sigma}\psi^{O}_{3,\sigma} $ & $\frac{g_{\nu}}{g_{\nu}^2+\cos\theta_{\nu}(g_{\nu}^2-1)+1} $\\ 

\hline \hline
\vspace{2mm}
Operator &  $\delta_{2\nu}$ \\
\hline
\vspace{2mm}
$\psi^{O~\dagger}_{1,\sigma}\psi^{I}_{1,\sigma}$    &
$\frac{1}{6}g_{\nu}(2-2\cos\theta_{\nu}) $\\

\vspace{2mm}
$\psi^{O~\dagger}_{2,\sigma}\psi^{I}_{2,\sigma}$   &
$\frac{1}{6}g_{\nu}(2+\cos\theta_{\nu}+\sqrt{3}\sin\theta_{\nu})$\\ 

\vspace{2mm}
$\psi^{O~\dagger}_{3,\sigma}\psi^{I}_{3,\sigma}$   &
$\frac{1}{6}g_{\nu}(2+\cos\theta_{\nu}-\sqrt{3}\sin\theta_{\nu}) $\\ 

\vspace{2mm}
$\psi^{O~\dagger}_{1,\sigma}\psi^{I}_{2,\sigma},\psi^{O~\dagger}_{2,\sigma}\psi^{I}_{1,\sigma}$ & $\frac{3+g_{\nu}^2}{24g_{\nu}}(2-2\cos\theta_{\nu}) $\\

\vspace{2mm}
$\psi^{O~\dagger}_{2,\sigma}\psi^{I}_{3,\sigma},\psi^{O~\dagger}_{3,\sigma}\psi^{I}_{2,\sigma}$  & $\frac{3+g_{\nu}^2}{24g_{\nu}}(2+\cos\theta_{\nu}-\sqrt{3}\sin\theta_{\nu}) $\\

\vspace{2mm}
$\psi^{O~\dagger}_{3,\sigma}\psi^{I}_{1,\sigma},\psi^{O~\dagger}_{1,\sigma}\psi^{I}_{3,\sigma}$  & $\frac{3+g_{\nu}^2}{24g_{\nu}}(2+\cos\theta_{\nu}+\sqrt{3}\sin\theta_{\nu}) $\\

\vspace{2mm}
$\psi^{O~\dagger}_{1,\sigma}\psi^{O}_{2,\sigma},\psi^{I~\dagger}_{1,\sigma}\psi^{I}_{2,\sigma}$  & $\frac{(g_{\nu}^2+1)+2(g_{\nu}^2-1)\cos\theta_{\nu}}{4g_{\nu}} $  \\

\vspace{2mm}
$\psi^{O~\dagger}_{2,\sigma}\psi^{O}_{3,\sigma},\psi^{I~\dagger}_{2,\sigma}\psi^{I}_{3,\sigma}$  & $\frac{(g_{\nu}^2+1)-(g_{\nu}^2-1)(\cos\theta_{\nu}-\sqrt{3}\sin\theta_{\nu})}{4g_{\nu}} $ \\

\vspace{2mm}
$\psi^{O~\dagger}_{3,\sigma}\psi^{O}_{1,\sigma},\psi^{I~\dagger}_{3,\sigma}\psi^{I}_{1,\sigma}$  & $\frac{(g_{\nu}^2+1)-(g_{\nu}^2-1)(\cos\theta_{\nu}+\sqrt{3}\sin\theta_{\nu})}{4g_{\nu}} $\\

\toprule
\end{tabular}

\end{table}

\appendix
\section{Scaling dimensions of the fixed points at the junction} 
\label{secIV}
\begin{table*}[h!]
\centering
\caption{Scaling dimension of various pair tunneling operators: $\delta_{0\sigma} = \delta_{kc} +  \delta_{ls}$, where $k,l = 1 {~\rm or~} 2$ and the boundary condition at the junction is specified by the matrix $\M_{kc} \M_{ls}$. \label{T2}}
\begin{tabular}{c c c c c}\toprule

\vspace{3mm}
Operator &  $\delta_{1c}$  & $\delta_{1s}$ & $\delta_{2c}$  & $\delta_{2s}$\\ 
\hline
%
$\psi^{O~\dagger}_{2,\uparrow}\psi^{I}_{1,\uparrow}\psi^{O~\dagger}_{2,\downarrow}\psi^{I}_{1,\downarrow}$    & 
$\frac{4g_c(2+\cos\theta_c+\sqrt{3}\sin\theta_c)}{3(1+g_c^2+(g_c^{2}-1)\cos\theta_c)}$   &0 & 
$\frac{3+g_c^2}{3g_c}(1-\cos\theta_c)$   & 0  \\
$\psi^{O~\dagger}_{3,\uparrow}\psi^{I}_{2,\uparrow}\psi^{O~\dagger}_{3,\downarrow}\psi^{I}_{2,\downarrow}$   &
"  & 0    &
$\frac{3+g_c^2}{6g_c}(2+\cos\theta_c-\sqrt{3}\sin\theta_c)$     & 0   \\
\vspace{3mm}
$\psi^{O~\dagger}_{1,\uparrow}\psi^{I}_{3,\uparrow}\psi^{O~\dagger}_{1,\downarrow}\psi^{I}_{3,\downarrow}$    &
"   & 0   &
$\frac{3+g_c^2}{6g_c}(2+\cos\theta_c+\sqrt{3}\sin\theta_c) $    &0 \\
%
$\psi^{O~\dagger}_{3,\uparrow}\psi^{I}_{1,\uparrow}\psi^{O~\dagger}_{3,\downarrow}\psi^{I}_{1,\downarrow}$   &
$\frac{4g_c(2+\cos\theta_c-\sqrt{3}\sin\theta_c)}{3(1+g_c^2+(g_c^{2}-1)\cos\theta_c)}$   & 0  &
$\frac{3+g_c^2}{6g_c}(2+\cos\theta_c+\sqrt{3}\sin\theta_c)$   & 0 \\
$\psi^{O~\dagger}_{1,\uparrow}\psi^{I}_{2,\uparrow}\psi^{O~\dagger}_{1,\downarrow}\psi^{I}_{2,\downarrow}$   &
"    & 0  & $\frac{3+g_c^2}{3g_c}(1-\cos\theta_c)$    & 0 \\
\vspace{3mm}
$\psi^{O~\dagger}_{2,\uparrow}\psi^{I}_{3,\uparrow}\psi^{O~\dagger}_{2,\downarrow}\psi^{I}_{3,\downarrow}$   &
"   & 0   & $\frac{3+g_c^2}{6g_c}(2+\cos\theta_c-\sqrt{3}\sin\theta_c)$    & 0\\
%
$\psi^{I~\dagger}_{2,\sigma}\psi^{I}_{1,\sigma}\psi^{O~\dagger}_{2,\sigma}\psi^{O}_{1,\sigma}$    &
$\frac{2g_c(1+\cos\theta_c)}{1+g_c^2+(g_c^{2}-1)\cos\theta_c}$  & $\frac{2g_s(1+\cos\theta_s)}{1+g_s^2+(g_s^{2}-1)\cos\theta_s}$   & 
$\frac{9(1-\cos\theta_c)-\sqrt{3}\sin\theta_c(1+2\cos\theta_c)}{9g_c}$    & 
$\frac{9(1-\cos\theta_s)-\sqrt{3}\sin\theta_s(1+2\cos\theta_s)}{9g_s}$   \\
$\psi^{I~\dagger}_{3,\sigma}\psi^{I}_{2,\sigma}\psi^{O~\dagger}_{3,\sigma}\psi^{O}_{2,\sigma}$   &
"    & "    &
$\frac{9(2+\cos\theta_c)-\sqrt{3}\sin\theta_c(11+4\cos\theta_c)}{18g_c}$  &
$\frac{9(2+\cos\theta_s)-\sqrt{3}\sin\theta_s(11+4\cos\theta_s)}{18g_s}$\\
\vspace{3mm}
$\psi^{I~\dagger}_{1,\sigma}\psi^{I}_{3,\sigma}\psi^{O~\dagger}_{1,\sigma}\psi^{O}_{3,\sigma}$    &
"   & "   &
$\frac{9(2+\cos\theta_c)+\sqrt{3}\sin\theta_c(7-4\cos\theta_c)}{18g_c}$   &
$\frac{9(2+\cos\theta_s)+\sqrt{3}\sin\theta_s(7-4\cos\theta_s)}{18g_s}$\\
%
$\psi^{I~\dagger}_{2,\sigma}\psi^{I}_{1,\sigma}\psi^{O~\dagger}_{2,-\sigma}\psi^{O}_{1,-\sigma}$ &
$\frac{2g_c(1+\cos\theta_c)}{1+g_c^2+(g_c^{2}-1)\cos\theta_c}$   & $\frac{2g_s(1-\cos\theta_s)}{1+g_s^2+(g_s^{2}-1)\cos\theta_s}$ &
$\frac{9(1-\cos\theta_c)-\sqrt{3}\sin\theta_c(1+2\cos\theta_c)}{9g_c}$ &
$\frac{g_s(9(1+\cos\theta_s)-\sqrt{3}\sin\theta_s(1+2\cos\theta_s))}{9}$\\
$\psi^{I~\dagger}_{3,\sigma}\psi^{I}_{2,\sigma}\psi^{O~\dagger}_{3,-\sigma}\psi^{O}_{2,-\sigma}$ &
"   & "   &
$\frac{9(2+\cos\theta_c)-\sqrt{3}\sin\theta_c(11+4\cos\theta_c)}{18g_c}$ &
$\frac{g_s(9(2-\cos\theta_s)+\sqrt{3}\sin\theta_s(7-4\cos\theta_s))}{18}$\\
\vspace{3mm}
$\psi^{I~\dagger}_{1,\sigma}\psi^{I}_{3,\sigma}\psi^{O~\dagger}_{1,-\sigma}\psi^{O}_{3,-\sigma}$  &
"   & "   &
$\frac{9(2+\cos\theta_c)+\sqrt{3}\sin\theta_c(7-4\cos\theta_c)}{18g_c}$&
$\frac{g_s(9(2-\cos\theta_s)-\sqrt{3}\sin\theta_s(11+4\cos\theta_s))}{18}$\\
%
$\psi^{O~\dagger}_{1,\uparrow}\psi^{I}_{1,\uparrow}\psi^{O~\dagger}_{1,\downarrow}\psi^{I}_{1,\downarrow}$ &
$\frac{8g_c(1-\cos\theta_c)}{3(1+g_c^2+(g_c^{2}-1)\cos\theta_c)}$& 0 &
$\frac{2g_c}{3}(2+\cos\theta_c-\sqrt{3}\sin\theta_c)$& 0\\
$\psi^{O~\dagger}_{2,\uparrow}\psi^{I}_{2,\uparrow}\psi^{O~\dagger}_{2,\downarrow}\psi^{I}_{2,\downarrow}$ &
" & 0 &
$\frac{(2g_c)}{3}(2+\cos\theta_c+\sqrt{3}\sin\theta_c)$ &0 \\
\vspace{3mm}
$\psi^{O~\dagger}_{3,\uparrow}\psi^{I}_{3,\uparrow}\psi^{O~\dagger}_{3,\downarrow}\psi^{I}_{3,\downarrow}$ &
"& 0 & $\frac{(4g_c)}{3}(1-\cos\theta_c)$ & 0 \\
%
$\psi^{O~\dagger}_{1,\sigma}\psi^{I}_{1,\sigma}\psi^{O~\dagger}_{2,-\sigma}\psi^{I}_{2,-\sigma}$ &
$\frac{2g_c(1-\cos\theta_c)}{3(1+g_c^2+(g_c^{2}-1)\cos\theta_c)}$& $\frac{2g_s(1-\cos\theta_s)}{(1+g_s^2+(g_s^{2}-1)\cos\theta_s)}$ &
$\frac{g_c}{3}(1-\cos\theta_c)$ & $g_s(1+\cos\theta_s)$\\
$\psi^{O~\dagger}_{2,\sigma}\psi^{I}_{2,\sigma}\psi^{O~\dagger}_{3,-\sigma}\psi^{I}_{3,-\sigma}$  &
" & " &
$\frac{g_c}{6}(2+\cos\theta_c-\sqrt{3}\sin\theta_c)$& $\frac{g_s}{2}(2-\cos\theta_s+\sqrt{3}\sin\theta_s)$\\
\vspace{3mm}
$\psi^{O~\dagger}_{3,\sigma}\psi^{I}_{3,\sigma}\psi^{O~\dagger}_{1,-\sigma}\psi^{I}_{1,-\sigma}$ &
" &"& 
$\frac{g_c}{6}(2+\cos\theta_c+\sqrt{3}\sin\theta_c)$ & $\frac{g_s}{2}(2-\cos\theta_s-\sqrt{3}\sin\theta_s)$\\
%
$\psi^{O~\dagger}_{1,\sigma}\psi^{I}_{1,\sigma}\psi^{O~\dagger}_{2,\sigma}\psi^{I}_{2,\sigma}$  &
$\frac{2g_c(1-\cos\theta_c)}{3(1+g_c^2+(g_c^{2}-1)\cos\theta_c)}$ & $\frac{2g_s(1-\cos\theta_s)}{3(1+g_s^2+(g_s^{2}-1)\cos\theta_s)}$ & $\frac{g_c}{3}(1-\cos\theta_c)$ & $\frac{g_s}{3}(1-\cos\theta_s)$\\
$\psi^{O~\dagger}_{2,\sigma}\psi^{I}_{2,\sigma}\psi^{O~\dagger}_{3,\sigma}\psi^{I}_{3,\sigma}$   &
" &" & $\frac{g_c}{6}(2+\cos\theta_c-\sqrt{3}\sin\theta_c)$ & $\frac{g_s}{6}(2+\cos\theta_s-\sqrt{3}\sin\theta_s)$\\
\vspace{3mm}
$\psi^{O~\dagger}_{3,\sigma}\psi^{I}_{3,\sigma}\psi^{O~\dagger}_{1,\sigma}\psi^{I}_{1,\sigma}$  &
" & "& $\frac{g_c}{6}(2+\cos\theta_c+\sqrt{3}\sin\theta_c)$ &$\frac{g_s}{6}(2+\cos\theta_s+\sqrt{3}\sin\theta_s)$\\
%
$\psi^{O~\dagger}_{2,\sigma}\psi^{I}_{1,\sigma}\psi^{O~\dagger}_{1,-\sigma}\psi^{I}_{2,-\sigma}$ &
$\frac{2g_c(1-\cos\theta_c)}{3(1+g_c^2+(g_c^{2}-1)\cos\theta_c)}$ & $\frac{2g_s(1+\cos\theta_s)}{(1+g_s^2+(g_s^{2}-1)\cos\theta_s)}$\ &
$\frac{g_c}{3}(1-\cos\theta_c)$ & $\frac{1}{g_s}(1-\cos\theta_s)$\\
$\psi^{O~\dagger}_{3,\sigma}\psi^{I}_{2,\sigma}\psi^{O~\dagger}_{2,-\sigma}\psi^{I}_{3,-\sigma}$ &
" & " &
$\frac{g_c}{6}(2+\cos\theta_c-\sqrt{3}\sin\theta_c)$ & $\frac{1}{2g_s}(2+\cos\theta_s-\sqrt{3}\sin\theta_s)$\\
\vspace{3mm}
$\psi^{O~\dagger}_{1,\sigma}\psi^{I}_{3,\sigma}\psi^{O~\dagger}_{3,-\sigma}\psi^{I}_{1,-\sigma}$ &
" &" & $\frac{g_c}{6}(2+\cos\theta_c+\sqrt{3}\sin\theta_c)$& $\frac{1}{2g_s}(2+\cos\theta_s+\sqrt{3}\sin\theta_s)$\\
%
$\psi^{O~\dagger}_{2,\uparrow}\psi^{I}_{1,\uparrow}\psi^{O}_{2,\downarrow}\psi^{I~\dagger}_{1,\downarrow}$ &
0  & $\frac{4g_s(2+\cos\theta_s+\sqrt{3}\sin\theta_s)}{3(1+g_s^2+(g_s^{2}-1)\cos\theta_s)}$ &
0& $\frac{3+g_s^2}{3g_s}(1-\cos\theta_s)$\\
$\psi^{O~\dagger}_{3,\uparrow}\psi^{I}_{2,\uparrow}\psi^{O}_{3,\downarrow}\psi^{I~\dagger}_{2,\downarrow}$  &
0 &" & 0 & $\frac{3+g_s^2}{6g_s}(2+\cos\theta_s-\sqrt{3}\sin\theta_s)$\\
\vspace{3mm}
$\psi^{O~\dagger}_{1,\uparrow}\psi^{I}_{3,\uparrow}\psi^{O}_{1,\downarrow}\psi^{I~\dagger}_{3,\downarrow}$ &
 0 & " & 0 & $\frac{3+g_s^2}{6g_s}(2+\cos\theta_s+\sqrt{3}\sin\theta_s)$\\
%
$\psi^{O~\dagger}_{3,\uparrow}\psi^{I}_{1,\uparrow}\psi^{O}_{3,\downarrow}\psi^{I~\dagger}_{1,\downarrow}$ &
0 & $\frac{4g_s(2+\cos\theta_s-\sqrt{3}\sin\theta_s)}{3(1+g_s^2+(g_s^{2}-1)\cos\theta_s)}$ &
0& $\frac{3+g_s^2}{6g_s}(2+\cos\theta_s+\sqrt{3}\sin\theta_s)$\\
$\psi^{O~\dagger}_{1,\uparrow}\psi^{I}_{2,\uparrow}\psi^{O}_{1,\downarrow}\psi^{I~\dagger}_{2,\downarrow}$ &
0 & " & 0 &  $\frac{3+g_s^2}{3g_s}(1-\cos\theta_s)$\\
\vspace{3mm}
$\psi^{O~\dagger}_{2,\uparrow}\psi^{I}_{3,\uparrow}\psi^{O}_{2,\downarrow}\psi^{I~\dagger}_{3,\downarrow}$  &
0 & " & 0 & $\frac{3+g_s^2}{6g_s}(2+\cos\theta_c-\sqrt{3}\sin\theta_s)$\\
%
$\psi^{O~\dagger}_{2,\sigma}\psi^{I}_{1,\sigma}\psi^{O}_{1,-\sigma}\psi^{I~\dagger}_{2,-\sigma}$ &
$\frac{2g_c(1+\cos\theta_c)}{1+g_c^2+(g_c^{2}-1)\cos\theta_c}$ & $\frac{2g_s(1-\cos\theta_s)}{3(1+g_s^2+(g_s^{2}-1)\cos\theta_s)}$ &
$\frac{1}{g_c}(1-\cos\theta_c)$& $\frac{g_s}{3}(1-\cos\theta_s)$\\
$\psi^{O~\dagger}_{3,\sigma}\psi^{I}_{2,\sigma}\psi^{O}_{2,-\sigma}\psi^{I~\dagger}_{3,-\sigma}$ &
 " &" & $\frac{1}{2g_c}(2+\cos\theta_c-\sqrt{3}\sin\theta_c)$ & $\frac{g_s}{6}(2+\cos\theta_s-\sqrt{3}\sin\theta_s)$\\
\vspace{3mm}
$\psi^{O~\dagger}_{1,\sigma}\psi^{I}_{3,\sigma}\psi^{O}_{3,-\sigma}\psi^{I~\dagger}_{1,-\sigma}$ &
" & " &
$\frac{1}{2g_c}(2+\cos\theta_c+\sqrt{3}\sin\theta_c)$& $\frac{g_s}{6}(2+\cos\theta_s+\sqrt{3}\sin\theta_s)$\\
\toprule
\end{tabular}
\end{table*}

In this appendix, we present the scaling dimensions of all the spin-preserving single particle and pair tunneling operators, for all possible fixed points.  For any boundary operator $\mathbb{O}_B$, the scaling dimension $\delta_0$ can be calculated by the two point correlation function
\be 
\langle \mathbb{O}_B(t)\mathbb{O}_B(t^{\prime}) \rangle \sim |t-t^{\prime}|^{-\delta_0}~,
\ee
and it depends on the strength of the e-e interactions ($g_\nu$), and the boundary condition at at the junction ($\theta_\nu$).
The stable boundary condition or fixed points are those for which the scaling dimension of all boundary operators is either $\delta_0=0$ or $\delta_0>1$ for a given e-e interaction strength. 
For a fixed point specified by $\M_{kc} \M_{ls}$, where $k, l = 1~{\rm or}~2$,  the scaling dimension of all single particle  and pair tunneling operators can be expressed as a sum of the charge and spin components: $\delta_0 = \delta_{kc} + \delta_{ls}$. For the single particle tunneling operators, $\delta_{1 \nu}$ is explicitly given in the upper half of Table~\ref{T1}, and $\delta_{2 \nu}$ is tabulated in the lower half of Table~\ref{T1}. The scaling dimensions of all possible pair tunneling operators, is tabulated in Table~\ref{T2}. 
Finally, we note that, ideally we need to know the scaling dimensions of all possible multi particle tunneling operators to determine the stability of a fixed point, however, more particle processes tend to be less relevant, and based on a conformal field theory argument it has been explicitly shown in Ref.~[\onlinecite{Hou}] that single and two particle tunneling operators completely determine the stability of all the fixed points which have a unique basin of attraction for a spin-1/2 Y-junction.


\begin{thebibliography}{77} 

\bibitem{Tomonaga} S. Tomonaga, 
\href{http://dx.doi.org/10.1143/ptp/5.4.544}{Prog. Theor. Phys. {\bf 5}, 544 (1950)}.

\bibitem{Luttinger} J. M. Luttinger, 
\href{http://dx.doi.org/10.1063/1.1704046}{J. Math. Phys. {\bf4}, 1154 (1963)}. 
%
\bibitem{Mattis} D. C. Mattis and E. H. Lieb, \href{http://dx.doi.org/10.1063/1.1704281}{J. Math. Phys. {\bf 6}, 304 (1965)}.
%
\bibitem{Haldane} F. D. Haldane, \href{http://dx.doi.org/10.1088/0022-3719/14/19/010}{J. Phys. C {\bf14}, 2585 (1981)}.
%
\bibitem{delft} J. V. Delft and H. Schoeller, \href{http://onlinelibrary.wiley.com/doi/10.1002/(SICI)1521-3889(199811)7:4\%3C225::AID-ANDP225\%3E3.0.CO;2-L/abstract}{Annalen Phys. {\bf7}, 225
(1998)}; S. Rao and D. Sen, in {\it Field Theories in Condensed
Matter Physics}, edited by S. Rao (Hindustan Book Agency, New Delhi, 2001). 

\bibitem{Giamarchi}T. Giamarchi, {\it Quantum Physics in One Dimension} (Oxford
University Press, Oxford, 2004).

\bibitem{Giovanni} D. Pines and P. Noz\"ieres, {\it The Theory of Quantum Liquids}
(W.A. Benjamin, Inc., New York, 1966); G.F. Giuliani and G. Vignale, {\it 
Quantum Theory of the Electron Liquid} (Cambridge University Press, Cambridge,
2005). 

\bibitem{Auslaender1}
O. M. Auslaender, H. Steinberg, A. Yacoby, Y. Tserkovnyak, B. I. Halperin, K. W. Baldwin, L. N. Pfeiffer, and K. W. West, 
\href{http://dx.doi.org/10.1126/science.1107821}{Science {\bf 308}, 88 (2005)}.
%
\bibitem{Jompol} Y. Jompol, C. J. B. Ford, J. P. Griffiths, I. Farrer, G. A. C. Jones, D. Anderson, D. A. Ritchie, T. W. Silk, and A. J. Schofield,
\href{http://dx.doi.org/10.1126/science.1171769}{Science {\bf 325}, 597 (2009)}.
%
\bibitem{Safi} I. Safi and H. J. Schulz, \href{http://dx.doi.org/10.1103/PhysRevB.52.R17040}{\prb ~{\bf 52}, R17040 (1995)}.
\bibitem{Safi2} I. Safi, 
\href{	http://dx.doi.org/10.1051/anphys:199705001}{Ann. Phys. (France) {\bf 22}, 463 (1997)}; \href{http://arxiv.org/abs/0906.2363}{arXiv: 0906.2363.} 

\bibitem{Safi3} I. Safi and H. J. Schulz, {\it Transport through a single-band channel 
connected to measuring leads} in Quantum Transport in Semiconductor 
Submicron Structures, edited by B. Kramer (Kluwer Academic Press, 
Dordrecht, 1995), Chap. 3, p. 159; {\it Transport in an interacting wire 
connected to measuring leads and proximity effects in correlated 
Fermions and Transport in Mesoscopic Systems}, edited by T. Martin, G. 
Montambaux, and J. T. T. Van (Editions Frontières, Gif-sur-Yvette, 
1996).


\bibitem{pham} K.-V. Pham, M. Gabay, and P. Lederer, \href{http://dx.doi.org/10.1007/s100510050799}{Eur. Phys. J. B {\bf 9}, 573 (1999)}; \href{http://dx.doi.org/10.1103/PhysRevB.61.16397}{Phys. Rev. B {\bf 61}, 16397 (2000)}.
\bibitem{Steinberg} H. Steinberg, 
G. Barak, A. Yacoby, L. N. Pfeiffer, K. W. West, B. I. Halperin, and K. Le Hur,
\href{http://dx.doi.org/10.1038/nphys810}{Nature Phys. {\bf 4}, 116 (2008)}.
\bibitem{Kamata} H. Kamata,	 N. Kumada,	 M. Hashisaka,	 K. Muraki, and T. Fujisawa, \href{http://dx.doi.org/10.1038/nnano.2013.312}{Nature Nanotechnology {\bf 9}, 177 (2014)}.

\bibitem{Chang} A. M. Chang, \href{http://dx.doi.org/10.1103/RevModPhys.75.1449}{Rev. Mod. Phys. {\bf 75}, 1449 (2003)}.

\bibitem{Bockrath} M. Bockrath, D. H. Cobden, J. Lu, A. G. Rinzler, R. E. Smalley, L. Balents and P. L. McEuen, \href{http://dx.doi.org/10.1038/17569}{Nature {\bf 397}, 598 (1999)}.

\bibitem{Jezouin}
S. Jezouin, M. Albert, F. D. Parmentier, A. Anthore, U. Gennser,	 A. Cavanna, I. Safi,	and F. Pierre, 
\href{http://dx.doi.org/10.1038/ncomms2810}{Nature Communications {\bf 4}, 1802 (2013)}.

\bibitem{Blumenstein} C. Blumenstein,	J. Sch\"afer,	S. Mietke,	S. Meyer,	A. Dollinger, M. Lochner, X. Y. Cui, L. Patthey, R. Matzdorf	 and R. Claessen, \href{http://dx.doi.org/10.1038/nphys2051}{Nature Physics {\bf 7}, 776 (2011)}.  

\bibitem{flensberg}
H. Bruus, and K. Flensberg, {\it Many-body quantum theory in condensed matter physics: An introduction}, (Oxford University Press, 02-Sep-2004).

\bibitem{fuhrer} M. S. Fuhrer, J. Nygard, L. Shih, M. Forero, Y.-G. Yoon, 
M. S. C. Mazzoni, H. J. Choi, J. Ihm, S. G. Louie, A. Zettl, and P. L. McEuen, 
\href{http://dx.doi.org/10.1126/science.288.5465.494}{Science {\bf 288}, 
494 (2000)}.
%
\bibitem{terrones} M. Terrones, F. Banhart, N. Grobert, J.-C. Charlier, H. Terrones, and P. M. Ajayan, 
\href{http://dx.doi.org/10.1103/PhysRevLett.89.075505}{\prl ~ {\bf 89}, 
075505 (2002)}. 
%

\bibitem{nayak} C. Nayak, M. P. A. Fisher, A. W. W. Ludwig, and H. H. Lin, 
\href{http://dx.doi.org/10.1103/PhysRevB.59.15694}{\prb ~ {\bf 59}, 15694 
(1999)}.
%

\bibitem{lal200} S. Lal, S. Rao, and D. Sen, 
\href{http://dx.doi.org/10.1103/PhysRevB.66.165327}{\prb~{\bf 66}, 165327 
(2002)}; S. Das, S. Rao, and D. Sen, 
\href{http://dx.doi.org/10.1103/PhysRevB.70.085318}{\prb ~{\bf 70}, 
085318 (2004)}.
%

\bibitem{chen} S. Chen, B. Trauzettel, and R. Egger, 
\href{http://dx.doi.org/10.1103/PhysRevLett.89.226404}{\prl ~{\bf 89}, 226404 
(2002)}. R. Egger, B. Trauzettel, S. Chen, and F. Siano, 
\href{http://dx.doi.org/10.1088/1367-2630/5/1/117}{New Journal of 
Physics {\bf 5}, 117 (2003)}.
%

\bibitem{chamon1} C. Chamon, M. Oshikawa, and I. Affleck, 
\href{http://dx.doi.org/10.1103/PhysRevLett.91.206403}{\prl ~\ {\bf 91}, 
206403 (2003)}; M. Oshikawa, C. Chamon, and I. Affleck, 
\href{http://dx.doi.org/10.1088/1742-5468/2006/02/P02008}{
J. Stat. Mech., P02008 (2006)}. 

\bibitem{Meden_prb2005} X. Barnab\'e-Th\'eriault., A. Sedeki, V. Meden, and K. 
Sch\"onhammer, \href{http://dx.doi.org/10.1103/PhysRevB.71.205327}{Phys. Rev. 
B, {\bf71} 205327(2005)}; 
\href{http://dx.doi.org/10.1103/PhysRevLett.94.136405}{Phys. Rev. Lett., 
{\bf 94} 136405 (2005)}.

\bibitem{das2} S. Das, S. Rao, and D. Sen, \href{http://dx.doi.org/10.1103/PhysRevB.74.045322}{\prb~{\bf 74} , 045322 (2006)}.

\bibitem{Giuliano} D. Giuliano and P. Sodano, \href{http://dx.doi.org/10.1016/j.nuclphysb.2008.11.011}{Nucl. Phys. B, {\bf 811} 395 (2009)}; \href{http://dx.doi.org/10.1088/1367-2630/10/9/093023}{New J. Phys. {\bf 10}, 093023 (2008)}. 

\bibitem{Bellazzini} B. Bellazzini, M. Burrello, M. Mintchev, and P. Sorba, 
Proceedings of Symposia in Pure Mathematics, Vol. 77 (American Mathematical 
Society, Providence) 2008, p. 639; B. Bellazzini, P. Calabrese, and 
M. Mintchev, \href{http://dx.doi.org/10.1103/PhysRevB.79.085122}{Phys. 
Rev. B {\bf 79}, 085122 (2009)}.

\bibitem{Hou} C.-Y. Hou and C. Chamon, 
\href{http://dx.doi.org/10.1103/PhysRevB.77.155422}{Phys. Rev. B {\bf 77}, 
155422 (2008)}.

%
\bibitem{dasrao} S. Das and S. Rao, \href{http://dx.doi.org/10.1103/PhysRevB.78.205421}{\prb~{\bf78} , 205421 (2008)}. 

%
\bibitem{agarwal_tdos} A. Agarwal, S. Das, S. Rao, and D. Sen, \href{http://dx.doi.org/10.1103/PhysRevLett.103.026401}
{Phys. Rev. Lett. {\bf 103}, 026401 (2009)}; 079903(E).
%
\bibitem{agarwal_diss} A. Agarwal, S. Das, and D. Sen, \href{http://dx.doi.org/10.1103/PhysRevB.81.035324}{Phys. Rev. B {\bf 81}, 035324 (2010)}.
%
\bibitem{abhiram} A. Soori and D. Sen, \href{http://dx.doi.org/10.1209/0295-5075/93/57007}{EPL {\bf 93}, 57007 (2011)}; \href{http://dx.doi.org/10.1103/PhysRevB.84.035422}{Phys. Rev. B {\bf 84}, 035422 (2011)}.
%
\bibitem{Rahmani} A. Rahmani, C.-Y. Hou, A. Feiguin, C. Chamon, and I. Affleck, \href{http://dx.doi.org/10.1103/PhysRevLett.105.226803}{\prl {\bf 105}, 226803 (2010)}. A. Rahmani et. al., 
\href{http://dx.doi.org/10.1103/PhysRevB.85.045120}{\prb~{\bf 85}, 045120 (2012)}.
%
\bibitem{Feldman_PRB2011} C. Wang, and D. E. Feldman, \href{http://dx.doi.org/10.1103/PhysRevB.83.045302}{Phys. Rev. B {\bf 83}, 045302 (2011)}.
%
\bibitem{Hou_prb2012} C. Y. Hou, A. Rahmani, A. E. Feiguin, and C. Chamon, \href{http://dx.doi.org/10.1103/PhysRevB.86.075451}{Phys. Rev. B {\bf 86}, 075451 (2012)}.
%
\bibitem{wolfe1} D. N. Aristov and P. W\"olfle, 
\href{http://dx.doi.org/10.1103/PhysRevB.86.035137}{\prb~{\bf 86}, 
035137 (2012)};
D. N. Aristov and P. W\"olfle, 
\href{http://dx.doi.org/10.1103/PhysRevB.88.075131}{\prb~{\bf 88}, 
075131 (2013)}. 
%
\bibitem{Agarwal_TRP} A. Agarwal, \href{http://dx.doi.org/10.1103/PhysRevB.90.195403}{Phys. Rev. B {\bf 90}, 195403 (2014)}.
%
\bibitem{Eggert} 
S. Eggert, \href{http://dx.doi.org/10.1103/PhysRevLett.84.4413}{Phys. Rev. Lett. {\bf 84}, 4413 (2000)}; 
S. Eggert, H. Johannesson, and A. Mattsson, \href{http://dx.doi.org/10.1103/PhysRevLett.76.1505}{Phys. Rev. Lett. {\bf 76}, 1505 (1996)}. 
%

\bibitem{Anfuso} F. Anfuso, and S. Eggert, 
\href{http://dx.doi.org/10.1103/PhysRevB.68.241301}{Phys. Rev. B {\bf 68}, 241301(R) (2003)}. 

\bibitem{Guigou} M. Guigou, T. Martin, and A. Crepieux, 
\href{http://dx.doi.org/10.1103/PhysRevB.80.045420}{Phys. Rev. B {\bf 80}, 045420 (2009)}; 
\href{http://dx.doi.org/10.1103/PhysRevB.80.045421}{Phys. Rev. B 80, 045421 (2009)}.

\bibitem{Pugnetti} S. Pugnetti, F. Dolcini, D. Bercioux, and H. Grabert, 
\href{http://dx.doi.org/10.1103/PhysRevB.79.035121}{Phys. Rev. B {\bf 79}, 035121 (2009)}. 

\bibitem {Ziani} N. Traverso Ziani, F. Cavaliere, and M. Sassetti, 
\href{http://dx.doi.org/10.1209/0295-5075/102/47006}{Europhys. Lett. {\bf 102}, 47006 (2013)}. 

\bibitem{Dario} D. Bercioux, G. Buchs, H. Grabert, and O. Gr\"oning, 
\href{http://dx.doi.org/10.1103/PhysRevB.83.165439}{Phys. Rev. B {\bf 83}, 165439 (2011)}.  

\bibitem{Ziani2} N. Traverso Ziani, G. Piovano, F. Cavaliere, and M. Sassetti, 
\href{http://dx.doi.org/10.1103/PhysRevB.84.155423}{Phys. Rev. B {\bf 84}, 155423 (2011)}. 


\bibitem{Voit} J. Voit, \href{http://dx.doi.org/10.1088/0953-8984/5/44/020}{J. Phys.: Condens. Matter {\bf 5}, 8305 (1993)};
K. Sch\"onhammer and V. Meden, \href{http://dx.doi.org/10.1103/PhysRevB.47.16205}{Phys. Rev. B 47, 16 205 (1993)}.  
\bibitem{Oreg} Y. Oreg and Alexander M. Finkel'stein, \href{http://dx.doi.org/10.1103/PhysRevLett.76.4230}{Phys. Rev. Lett. {\bf 76}, 4230 (1996)}. 
%
\bibitem{Fabrizio} M. Fabrizio and A. O. Gogolin, \href{http://dx.doi.org/10.1103/PhysRevLett.78.4527}{Phys. Rev. Lett. {\bf 78}, 4527 (1997)}.
%
\bibitem{Furusaki} A. Furusaki, \href{http://dx.doi.org/10.1103/PhysRevB.56.9352}{Phys. Rev. B {\bf 56}, 9352 (1997)}; 
%
\bibitem{Braunecker} B. Braunecker, C. Bena, and P. Simon, \href{http://dx.doi.org/10.1103/PhysRevB.85.035136}{Phys. Rev. B. {\bf 85}, 035136 (2012)}.
\bibitem{Fazio} C. Winkelholz, R. Fazio, F. W. J. Hekking, and G. Sch{\"o}n, \href{http://dx.doi.org/10.1103/PhysRevLett.77.3200}{Phys. Rev. Lett. {\bf 77}, 3200 (1996)}. 
%
\bibitem{Liu} D. E. Liu and A. Levchenko, \href{http://dx.doi.org/10.1103/PhysRevB.88.155315}{Phys. Rev. B {\bf 88}, 155315 (2013)}.
%
\bibitem{CNTLL} C. Kane, L. Balents, and M. P. A. Fisher, \href{http://dx.doi.org/10.1103/PhysRevLett.79.5086}{Phys. Rev. Lett. {\bf 79}, 5086 (1997)}.

\end{thebibliography}
\end{document}